\documentclass[12pt]{article}

\usepackage{graphics} %   <------- for LatTeX + PS
\usepackage{epsfig}

\begin{document}
\parindent 0pt
\vskip 0.5 truecm

\parindent 0pt
\vskip 0.5 truecm
\begin{center}
\Large {\bf On Solvable Potentials for One Dimensional
Schr\"odinger and Fokker-Planck Equations}\\
\vspace{0.25in} \normalsize George Krylov \\

\vspace{0.3in}
Department of Physics, Belarusian State University,\\
Fr. Skariny av. 4, 220050 Minsk, Belarus\\
e--mail:  krylov@dragon.bas-net.by\\

\end{center}
\parskip 10pt

\vspace{0.3in}

\normalsize \noindent {\bf Abstract.} { One construction of
 exactly-solvable potentials for Fokker-Planck equation
 is considered based on supersymmetric quantum mechanics
 approach.}

\vspace{0.6in}

Submitted to {\bf Journal of Physics A: Mathematical and General}
\normalsize \vspace{0.1in}

\newpage

\section{Introduction}
Exactly solvable models play an important role in the developing of all branches of
physics as well as persist to be a genuine challenge to our understanding  of nature
beauty and simplicity. Comprehensive analysis of underlying reasons of solvability is
still a dream today even in such developed field as quantum mechanics, whereas solvable
models known in all other branches of physics, could be easily enumerate by hands.

Among the most developed approaches to solvability problem in
quantum mechanics is super\-symmetric quantum mechanics (SUSY QM)
approach that has much in common with the early Infeld \& Hull
factorization \cite{Infeld-Hull51}. It was developed  by E.~Witten
\cite{Witten81} originally in the context of quantum field theory,
but later gathered a lot of attention especially after
L.~Genderstein introduction of the concept of self-similar
potential \cite{Gendenstein83}. Today the SUSY QM approach has
clearly outlined the region of its applicability as well as its
connection with other analytical tools and methods. Some
interesting new results on the topic were obtained in the way of
generalization of SUSY approach on matrix models, higher order
symmetries and can be found in
\cite{Andrianov-et-al97}-\cite{Tkachuk99}).

The another type of approaches of the solvability problem have to be mentioned, which
were developed separately though been and becoming more and more closer to SUSY QM, are
numerous algebraic approaches. In this context we would like to mention only a few
papers, firstly  relatively old classical papers
\cite{Olshanetsky-Perelomov83},\cite{Alhassid-et-al83b} with extensive reference lists
therein and some recent ones
\cite{Kakei97}-%, \cite{Horozov-Kasman98}, \cite{Quesne99},
\cite{Beckers-et-al99} discussing different aspects of the problem.

There are also analytical ways of attacking the problem which are also very numerous and
we only cite some recent papers
\cite{Krahmer-et-al98}-%,\cite{Bender-et-al-99},
\cite{Robnik-Romanovski99},
\cite{Krylov-Robnik2000} demonstrating
some new ideas but been away out of  the topics discussed in the
present work.

For  Fokker-Planck equations solvable problems are still very rear
events as well as developed approaches to this problem. Some
recent results have been obtained for the generalization of SUSY
QM approach on this situation \cite{Junker98} (see also the
references therein).

The goal of the present paper is to establish close relations
between the above mentioned equations based on the SUSY approach
developed in \cite{Junker98}. As it will be demonstrated, the set
of exactly solvable potentials for the Fokker-Planck equation is
at least not poorer than that of Schr\"odinger one.

\section{Diffusion equation and Superpotential}
 The diffusion (Fokker-Planck) equation  for the distribution function
 $f(\mathbf{x},t)$ for a
 system in an external potential
 $U(\mathbf{x})$ has the following form
\begin{eqnarray} \label{dif-eq1}
 \frac{\partial f(\mathbf{x},t)}{\partial t}=
 \mathbf{\nabla}\cdot\hat\mathcal{D}\cdot\mathbf{\nabla}
 f(\mathbf{x},t) + \mathbf{\nabla}\left(f(\mathbf{x},t)\mathbf{\nabla}U(\mathbf{x})\right)
\end{eqnarray}
where $\mathbf{x}$ are some space co-ordinates for the system,
$\hat\mathcal{D}$ is a diffusion tensor and we assume in
subsequent that both diffusion tensor and external field potential
do not depend on time explicitly.

In literature, it is commonly accepted that the only difference of diffusion equation
(for spherically symmetric particles,when $\hat\mathcal{D}_{ij}=D\delta_{ij}$) and
Schr\"odinger's one is imaginary time on respect to real time (Vick's rotation). Though
it is evidently true for the case of a free particle, for the problem in an external
field the only sight  on the second equation
\begin{eqnarray}\label{schr-eq1}
 i\hbar\frac{\partial \psi(\mathbf{x},t)}{\partial t}=
 -\frac{\hbar^2}{2m}\mathbf{\nabla}^2 \psi(\mathbf{x},t) + U(\mathbf{x})\psi(\mathbf{x},t)
\end{eqnarray}
immediately demonstrates that the external field is incorporated
into the equation (\ref{schr-eq1}) in a different way on respect
to that in the diffusion case (\ref{dif-eq1}).

The prominent feature of the eq.(\ref{dif-eq1}) is the existence
of zero-mode (stationary or steady-state) solution
$f_s(\mathbf{x})$, which simply corresponds to generalization of
known Boltzmann distribution $f_s(\mathbf{x})=C\exp\left(-
U(\mathbf{x})\right)$ (valid when diffusion tensor is spherical).
Appropriate first order
 system of differential equations can be obtained from
eq.(\ref{dif-eq1}) as
\begin{eqnarray} \label{steady1} \nonumber
  \mathbf{\nabla}\cdot\hat\mathcal{D}\cdot\mathbf{\nabla}
 f_s(\mathbf{x}) +
 \mathbf{\nabla}\left(f_s(\mathbf{x})\mathbf{\nabla}U(\mathbf{x})\right)=
  \mathbf{\nabla}\left(\hat\mathcal{D}\cdot\mathbf{\nabla}
 f_s(\mathbf{x}) +
 f_s(\mathbf{x})\mathbf{\nabla}U(\mathbf{x})\right)=0
\end{eqnarray}
\begin{eqnarray} \label{steady2}
\hat\mathcal{D}\cdot\mathbf{\nabla}
 f_s(\mathbf{x}) +
 f_s(\mathbf{x})\mathbf{\nabla}U(\mathbf{x})= \mathbf{rot}\ \mathbf{b(x)}
\end{eqnarray}
where $\mathbf{b(x)}$ is some  vector function that should be
chosen in a way to allow a non-trivial solution of (\ref{steady2})
satisfying proper boundary conditions.

In opposite, for the Schr\"odinger equation (\ref{schr-eq1}) the ground state is
typically unknown and of most interest.

This, as we will see at least for one dimensional problems, is due
to the fact that after transforma\-tion of the diffusion equation
into the form of the Schr\"odinger one, we obtain the last in the
supersymmetric quantum mechanics ( SUSY) form directly and the
proper partner Hamiltonian is just $H_{-}$ \cite{Gendenstein83}.

Let us follow the way, similar to \cite{Junker98} and concentrate
in the subsequent  on 1D problems only (so that
$\mathbf{x}\rightarrow x$) . We also assume the  units' choice
such that $\hbar=1, m=1, D=1/2$. It is worth to note that steady
state solution  reads $f_s(x)=\exp(-2U(x))$  with this units'
choice.

Then, after substitution $f(x)=\exp\left\{-U(x)-Et\right\}\psi(x)$
into
\begin{eqnarray} \label{dif-eq2}
 \frac{\partial f(x,t)}{\partial t}=
 \frac{1}{2}\; \frac{\partial^2 f(x,t)}{\partial t^2} +
 \frac{\partial}{\partial x}\left(f(x,t)U'(x)\right)
\end{eqnarray}
we get the Schr\"odinger equation in the form
\begin{eqnarray} \label{to-schr1}
\frac{1}{2}\psi''(x) +\left(E - V_q (x)\right)\psi(x)=0
\end{eqnarray}
with a "quantum potential" $V_q(x)$ given
\begin{eqnarray}
V_q(x)= \frac{1}{2}\,{U'(x)}^2 -\frac{1}{2}\,U''(x)
\label{to-schr1a}
\end{eqnarray}

It is important to note that steady state solution of
(\ref{dif-eq2}) reads $f_0(x)=\exp(-2U(x))$ in our case

The last equation is just in the form of SUSY QM approach with the
{\it superpotential} given by $W(x)=U'(x)$ \cite{Junker98} and the
Hamiltonian operator having the factorized form
\begin{eqnarray} \label{to-schr2}
\hat H_{-}= \hat A^\dag \hat A
=\frac{1}{\sqrt{2}}\left(-\frac{d}{dx}+U'(x)\right)
\frac{1}{\sqrt{2}}\left(\frac{d}{dx}+U'(x)\right)
\end{eqnarray}
It is worth to remark that quantum potentials for  partner
Hamiltonians $\hat H_{\pm}$ \cite{Gendenstein83} now correspond
simply to $X$-axis reflection images of the original diffusion
potential $U(x)$.

It is immediately follows from the (\ref{to-schr1}) and
(\ref{to-schr2}) that the state $E=0$ is the eigenstate of
$H_{-}$.

The last clarifies the principal difference between the Schr\"odinger and  diffusion
equations. Whereas it can be highly nontrivial problem to construct an explicit
factorization of a given Hamiltonian and find as a result the ground state of a quantum
system, for a diffusion equation (at least for one-dimensional one) this is not a problem
at all, as it can be always done in the way been outlined.

To construct solvable cases for a 1-D diffusion equation, one can exploit the
supersymmetric form directly, considering known shape-invariant partner potentials
\cite{Junker98}.\\[0.2cm]

We will choose another way, namely we try to answer the following
question:
 \begin{quote}
  {\it what superpotential $W(x)=U'(x)$ should be that leads to
  exactly-solvable potentials for the eq.(\ref{to-schr1})}
 \end{quote}
provided we use most general form of a 1D quantum potential
allowing polynomial anzatz for a wave function
\cite{Krylov-Robnik2000}. \\[0.2cm]

Denoting a solvable quantum potential in (\ref{to-schr1}) by
$V_s(x)$, we consider now (\ref{to-schr1a}) as the Ricatti
equation for superpotential $W(x)$
\begin{eqnarray} \label{super1}
W'(x) - {W(x)}^2=-2\;V_s(x)
\end{eqnarray}

Here it worth to point out the following. We could split the
energy parameter $E$ in (\ref{to-schr1}) as $E=E_1+E_2$ that leads
to the appearance of one term e.g., $E_2$ in the right side of
(\ref{super1}), that will be used in subsequent.

Now, we  perform the known trick. Based on the correspondence of
Ricatti and Schr\"odinger equation we introduce substitution
\begin{equation}\label{substW}
W(x)=-\Psi'(x)/\Psi(x))
\end{equation}
and rewrite (\ref{super1}) in the form of the Schr\"odinger
equation for a function $\Psi(x)$

\begin{eqnarray}\label{eq:Sch1}
 \frac{1}{2}\Psi''(x) + \left(E_2  - V_s(x)\right)\Psi(x)=0
\end{eqnarray}\

The last simply means that every eigenstate  $\Psi_n(x)$ of a
quantum solvable potential $V_s(x)$ gives a superpotential through
the relation (\ref{substW}) that after integration gives for the
the diffusion equation potential $U(x)$ a simple formula

\begin{equation}\label{substW1}
U_n(x)=U_0+ \log \left |\Psi_n(x)\right|
\end{equation}

Most important fact here is that the set of $U_n(x)$ leads to the
same Schr\"odinger equation (\ref{to-schr1}) (with different
splitting of the constant $E$, but of course gives different
solutions for original diffusion equation (\ref{dif-eq1}).

 One more comment worth to be done is though representation in
the form () looks very known from the point of view of I Inverse
Scattering Transform theory (see, e.g. \cite{Ablowitz}) as well as
the formulae (\ref{super1},\ref{substW1}) are valid in all cases,
but only for solvable quantum potentials we obtain a closed form
solution of the original problem. Indeed, logarithm of some $n$-th
eigenfunction of a solvable potential $V_s(x)$ gives diffusion
potential (eq.(\ref{substW1})) whereas eigenfunctions modified by
exponential factor (written before eq.(\ref{dif-eq2})) give the
eigenfunctions of the diffusion problem itself.

Finally, we express the main result of the paper as follows. The
i-th eigenstate for some exactly solvable diffusion potential
$U_n(x)$ (given by above written formula ()) reads

\begin{equation}\label{substW2}
 f_i(x,t)=
\Psi_n(x) \exp\left(-(E_i+\Delta E) t\right) \Psi_i(x) \ \ \ \ \
i=0,1,...,n...
\end{equation}

where $E_i$ is eigenenergy of the appropriate quantum potential ()
and $\Delta E=-E_0$ is the constant energy shift introduce to give
 proper energy of the ground state ($E=0$ ) for the diffusion equation.

\section{Conclusion}
Simple construction we discussed opens a new sight on
interrelation of  Fokker-Planck and associate Schr\"odinger
problems. Having a significantly large number of exactly solvable
quantum potentials (see, e.g. list in \cite{Grosche-Steiner98})
one can construct at least measure similar list of solvable
diffusion cases having possibility to choose  one appropriate to
approximate some practical problems.

\section{Acknowledgement}
Author would like to thank Prof. Dr. Marko Robnik for numerous and
stimulating discussions of this and related problems. This work
has been support in parts by the Fund of Fundamental Researches of
the Republic of Belarus (Project F00-158), and Swiss Science
Foundation (Project SCOPES 7BYPJ065731).

\end{document}